\documentclass[superscriptaddress]{revtex4}

\usepackage{graphicx}
\newcommand{\gev}{\,\, \mathrm{GeV}}
\newcommand{\tev}{\,\, \mathrm{TeV}}
\newcommand{\mh}{m_h}
\newcommand{\mH}{m_H}
\newcommand{\tb}{\tan \beta}
\newcommand{\cp}{{\cal CP}}
\newcommand{\fh}{{\em FeynHiggs}}
\newcommand{\mt}{m_{t}}
\newcommand{\msq}{m_{\tilde{q}}}
\newcommand{\mgl}{m_{\tilde{g}}}
\def\aeff{\alpha_{\rm eff}}

\newcommand{\MA}{M_A}
\def\reffi#1{\mbox{Fig.~\ref{#1}}}
\def\refse#1{\mbox{Sect.~\ref{#1}}}
\newcommand{\mhmax}{m_h^{\rm max}}
\newcommand{\mste}{m_{\tilde{t}_1}}
\newcommand{\mstz}{m_{\tilde{t}_2}}
\newcommand{\msbe}{m_{\tilde{b}_1}}

\begin{document}
% You should use BibTeX and revtex.bst for references
\bibliographystyle{revtex}

\thispagestyle{empty}
\setcounter{page}{0}
\def\thefootnote{\fnsymbol{footnote}}

\begin{flushright}
BNL--HET--01/35\\
CALT-68-2353\\
DCPT/01/86\\IPPP/01/43\\
hep-ph/0110219\\
%\date{\today}
\end{flushright}

\vspace{1cm}

\begin{center}

{\Large\sc {\bf Soft SUSY-breaking scenarios \\in the light of Higgs 
searches at LEP2$^*$}}

\vspace{1cm}

{\sc A.~Dedes$^{\,1,\dag}$
S.~Heinemeyer$^{\,2,\ddag}$
S.~Su$^{\,3\S}$
G.~Weiglein$^{\,4\P}$
\vspace{0.5cm}
}

\vspace*{1cm}

$^1$ Physikalisches Institut der Universit\"at Bonn, 
 Nu\ss allee 12,  D-53115 Bonn, Germany

\vspace*{0.4cm}

$^2$ HET, Physics Department, Brookhaven Natl.\ Lab., Upton, NY
11973, USA

\vspace*{0.4cm}

$^3$ California Institute of Technology, Pasadena, CA 91125, USA

\vspace*{0.4cm}

$^4$ Institute for Particle Physics Phenomenology, University 
of Durham, Durham DH1~3LR, UK

\end{center}

\vspace*{1cm}

\begin{center}
{\large\bf Abstract}
\end{center}

  The Higgs boson sector of the Minimal Supersymmetric Standard Model (MSSM)
  is investigated in a uniform way in the framework of the three most 
  prominent soft SUSY-breaking scenarios, mSUGRA, mGMSB and mAMSB,
  especially concerning the Higgs searches at LEP2.
\vfill

$^*$ Contribution to the ``Workshop on the Future of
Particle Physics'' (snowmass 2001), Snowmass, USA, July 2001.

$^\dag$ dedes@th.physik.uni-bonn.de
 
$^\ddag$ Sven.Heinemeyer@bnl.gov

$^\S$ shufang@theory.caltech.edu

$^\P$ Georg.Weiglein@durham.ac.uk
\newpage

\section{Introduction}
\label{P3-05sec1}
The search for the light neutral Higgs boson is a crucial test of
Supersymmetry (SUSY) that can be performed with the present and the next
generation of high-energy colliders.
The data taken during the final year of LEP running at $\sqrt{s} \agt
206 \gev$, while establishing a 95\% C.L.\ exclusion limit for the Standard 
Model (SM) Higgs boson of $\mH > 114.1$~GeV, showed at about the
$2.1\sigma$ level an excess of signal-like events over the background 
expectation which is in agreement with the 
expectation for the production of a SM Higgs
boson of about 115~GeV~\cite{LEPHiggs}.
Within the MSSM, the LEP excess can be
interpreted as the production of the lightest $\cp$-even Higgs boson, which
over a wide parameter range has SM-like couplings, or of the heavier 
$\cp$-even Higgs boson, in a region of parameter space where the $\cp$-odd
Higgs boson $A$ is light and the ratio of the vacuum expectation values
of the two Higgs doublets, $\tb$, is relatively
large.  

In this work we investigate the predictions in the Higgs sector
arising from the three SUSY-breaking scenarios:  
minimal Supergravity (mSUGRA)~\cite{sugra}, 
minimal Gauge Mediated SUSY Breaking (mGMSB)~\cite{gmsb}
and minimal Anomaly Mediated SUSY Breaking 
(mAMSB)~\cite{amsb}. 
We relate the high energy input from these scenarios in a uniform way to the
predictions for the low-energy phenomenology in the Higgs sector,
allowing thus a direct comparison of the predictions arising from the
different scenarios. 
The MSSM Higgs masses and couplings are calculated using 
the program \fh~\cite{feynhiggs}. 
We analyze the consequences of the results obtained from the Higgs
search at LEP on the parameter space of the three scenarios.
For the case where LEP excess is interpreted as 
a possible Higgs signal, we furthermore discuss the
corresponding spectra of the SUSY particles in each scenario in view of 
the SUSY searches at the next generation of colliders.  For details of 
the calculations, see \cite{asbs1}.

\section{Input parameters and phenomenological restrictions}
\label{P3-05sec2}
For the numerical analysis we have scanned over about $50000$ 
models each for mSUGRA, mGMSB and mAMSB,
where the parameters have been randomly chosen in the intervals
as listed in Table~\ref{P3-05tab1}.
\begin{table}[b]
\caption{Input parameter ranges for mSUGRA, mGMSB and mAMSB, 
respectively. }
\label{P3-05tab1}
\begin{tabular}{rclrclrcl}\hline
\multicolumn{3}{c}{mSUGRA}&\multicolumn{3}{c}{mGMSB}&\multicolumn{3}{c}{mAMSB}
\\ \hline
$50~{\rm GeV} \le$&${M}_0$&$\le 1~{\rm TeV}$&
$\ \ \ 10^4 \gev \le$&$\Lambda$&$\le 2\,\times\,10^5 \gev\ \ \ $&
$20 \tev \le$&${m}_{\rm aux}$&$\le 100 \tev$ \\

$50~{\rm GeV} \le$&$M_{1/2}$&$\le 1~{\rm TeV}$&
$1.01\,\Lambda \le$&${M}_{\rm mess}$&$\le 10^5\,\Lambda$&
$0 \le$&$m_0$&$\le 2 \tev$\\

$-3~{\rm TeV} \le$&$A_0$&$\le 3~{\rm TeV}$&
$1 \le$&$N_{\rm mess}$&$\le 8$&\\

$1.5 \le$&$\tb$&$\le 60$&
$1.5 \le$&$\tb$&$\le 55$&
$1.5 \le$&$\tb$&$\le 60$ \\

&${\rm sign}\, \mu$&$= \pm 1$&
&${\rm sign}\, \mu$&$= \pm 1$&
&${\rm sign}\, \mu$&$= \pm 1$ \\ \hline
\end{tabular}
\end{table}

We also take into account some further
constraints when determining the allowed parameter values.
We require the contribution to the
$\rho$-parameter to be smaller than $3\times{10}^{-3}$~\cite{pdg}.
We impose the lower limits on the SUSY particle masses based on the 
negative search results of Run I of the Tevatron and at LEP~\cite{pdg}.
For the top-quark mass, throughout this paper we use the value
$\mt = 175 \gev$. A variation of $\mt$ by $\pm1 \gev$ would results in a
change in $\mh$ of about $\pm 1 \gev$~\cite{tbexcl}.
The GUT or high-energy scale parameters are taken to be real and 
$R$-parity symmetry is taken to be conserved.
We require successful radiative electroweak symmetry breaking (REWSB)
and parameter sets that do not fulfill the Charge-Color-Breaking 
constraints are discarded.  We have imposed relatively mild 
naturalness upper bounds:
$\msq \alt 1.5 \tev,\ \mgl \alt 2 \tev $.
We furthermore demand that the lightest SUSY particle is uncolored 
and uncharged.  On the other hand,
we do not demand a relic density in the region favored by 
dark matter constraints.  As a conservative approach, we do not 
apply any further constraints from $g_{\mu} -2$ or $b \to s \gamma$
(details can be found in~\cite{asbs1}).

For our numerical analysis we will focus on
three different cases implying different restrictions on the MSSM
parameter space. 
\begin{itemize}
\item[(I)]
We investigate the full parameter space
which is allowed in the three scenarios when taking into account the
exclusion bounds from the Higgs search \cite{LEPHiggs,mssmhiggs}
and the further constraints
discussed above. 
\item[(II)]
The LEP excess is interpreted as production of the 
lightest $\cp$-even Higgs boson of the MSSM: $\mh = 115 \pm 2 \gev$.
In order to allow this interpretation, $h$
has to have SM-like couplings to the $Z$, i.e.\ $\sin^2(\beta-\aeff) \agt 0.8$.
We also require that the decay of the light $\cp$-even
Higgs boson is SM like, i.e.\ the dominating decay channel is
$h\to{b}\bar b$. 
The $hb\bar b$ coupling is
mainly altered in two ways compared to the SM: it has an extra
factor $\sin\aeff/\cos\beta$ and it receives a correction 
$\sim 1/(1 + \Delta m_b)$. 
Therefore we demand $\sin^2\aeff/\cos^2\beta \agt 0.8$
and $|\Delta m_b| < 0.5$.
\item[(III)]
The LEP excess is interpreted as production of the 
heavy $\cp$-even Higgs boson of the MSSM: $\mH = 115 \pm 2 \gev$.
To have SM-like $ZZH$ and $Hb\bar{b}$ coupling, we require 
$\cos^2(\beta-\aeff) \agt 0.8$ and $\cos^2\aeff/\cos^2\beta \agt 0.8$.
In addition, we apply a bound of $\mh + \MA > 206 \gev$
as the associated production $e^+e^- \to Ah$, 
being $\sim \cos^2(\beta-\aeff)$, 
has to be beyond the kinematic
reach of LEP.
\end{itemize}

\section{Numerical analysis}
\begin{figure}[ht!]
\begin{center}
\includegraphics[width=5cm, height=4cm]{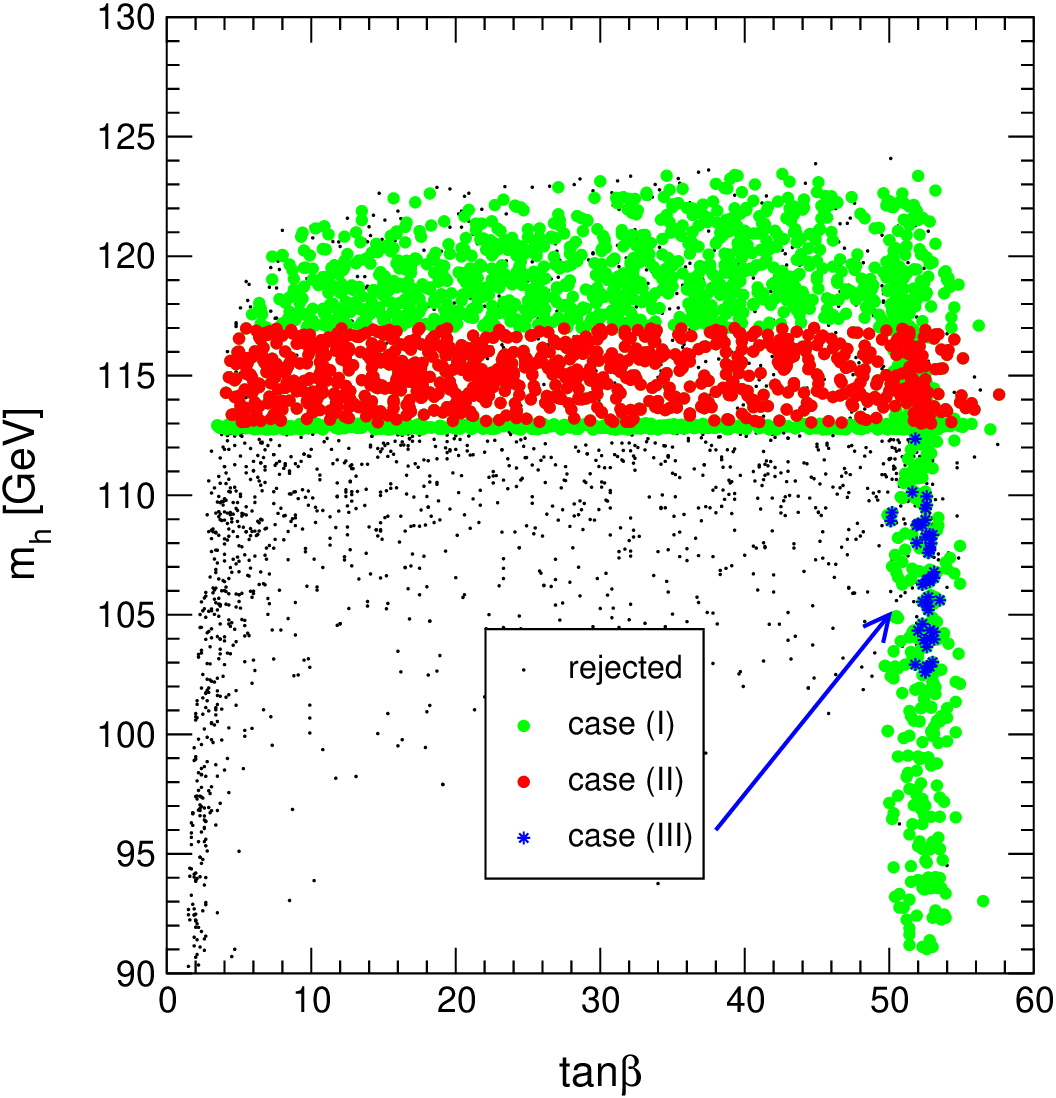}
\hspace{0.5cm}
\includegraphics[width=5cm, height=4cm]{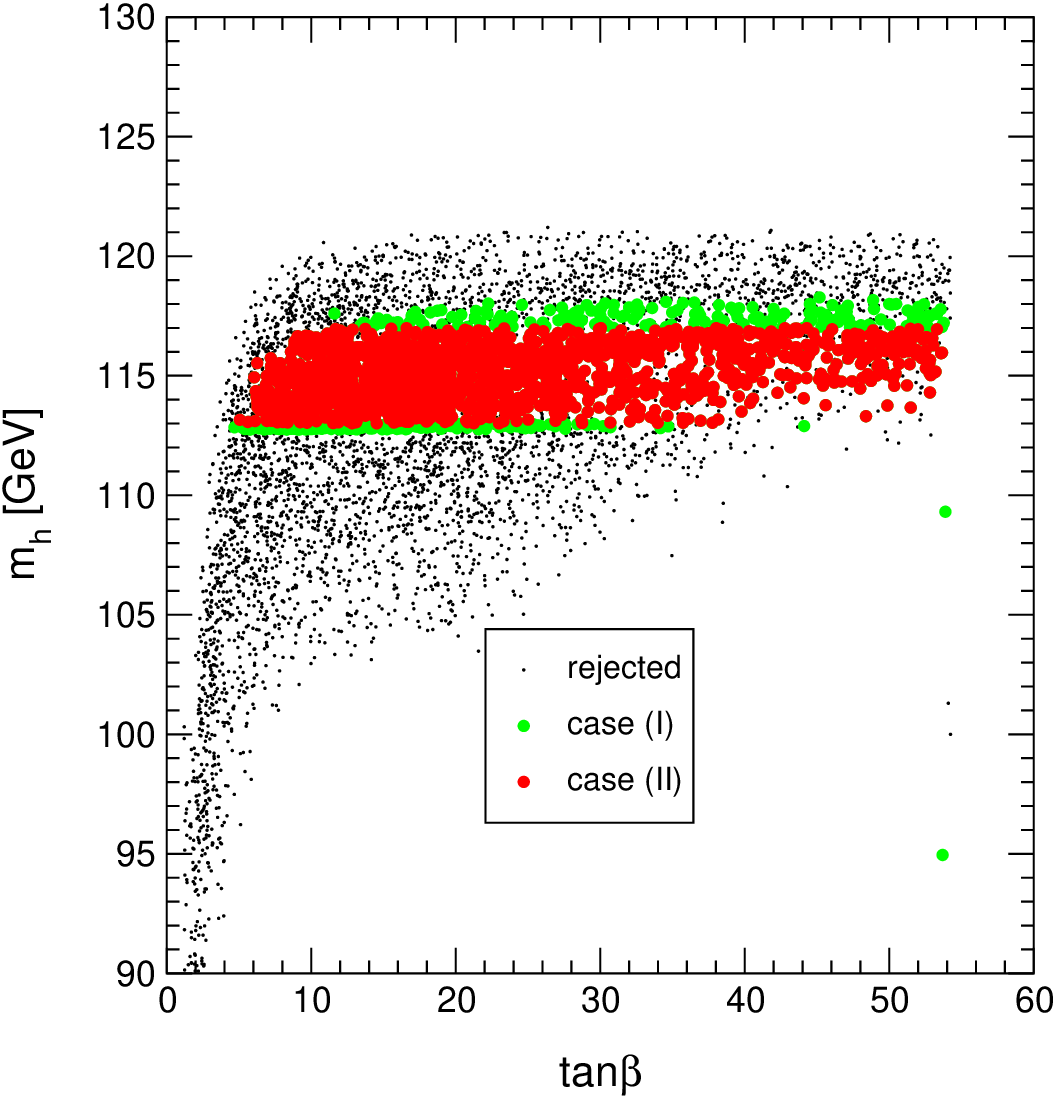}
\hspace{0.5cm}
\includegraphics[width=5cm, height=4cm]{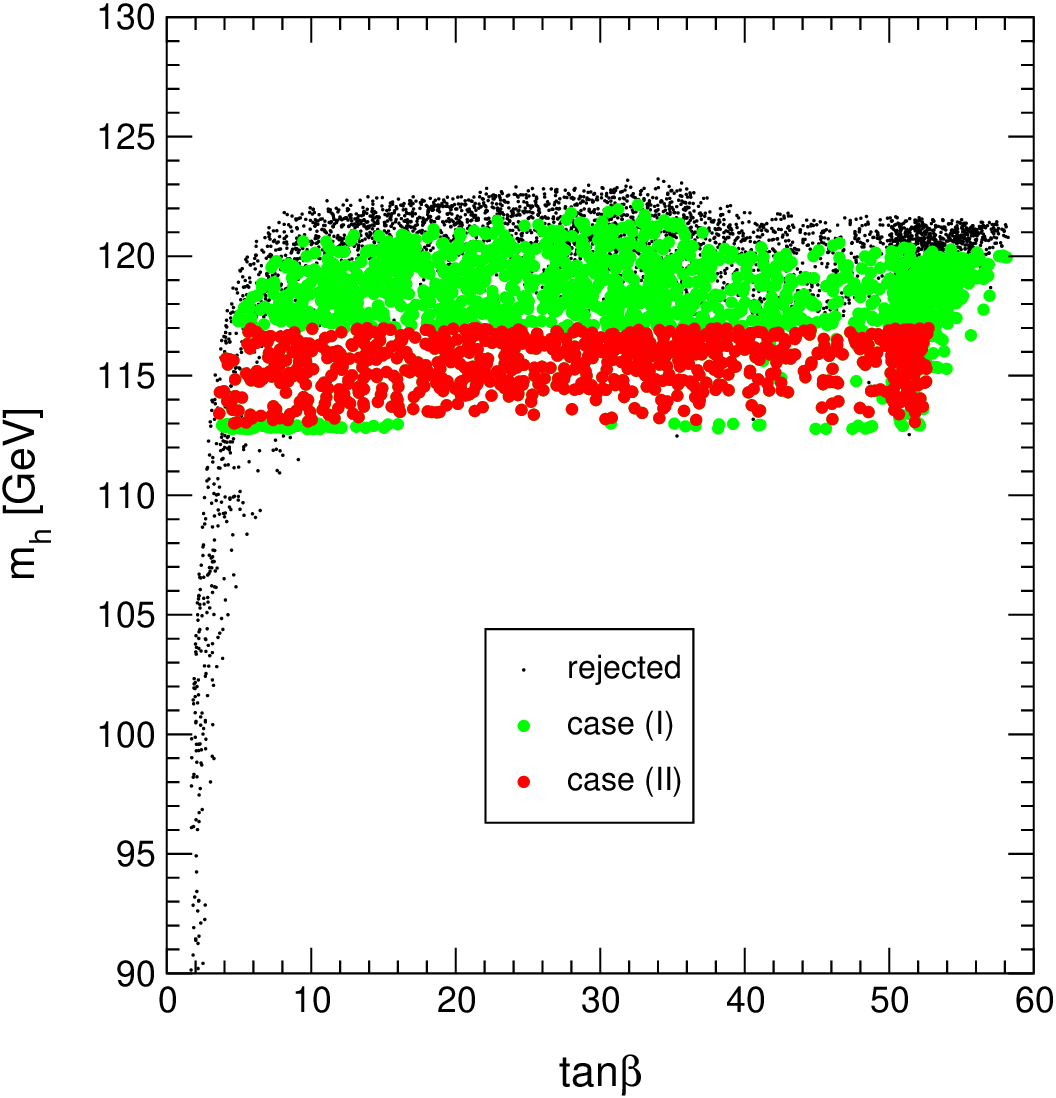}
\caption{
The light $\cp$-even Higgs boson mass $\mh$ as a function of 
$\tb$ in the mSUGRA (left plot), mGMSB (middle plot) and mAMSB (right plot) 
scenarios. }
\label{P3-05fig1}
\end{center}
\end{figure}
In \reffi{P3-05fig1} we show the variation of the 
light Higgs boson mass with respect to $\tb$ 
for the three cases defined in \refse{P3-05sec2}. 
$\mh$ sharply increases with $\tb$ in the region
of low $\tb$, while for $\tb \agt 10$ the $\mh$ values saturate. 
Values of $\tb \agt 60$ are not allowed due to the REWSB constraint.
The LEP2 Higgs boson searches 
exclude the models with $\mh \alt 113 \gev$ for $\tb \alt 50$.
Case (III) can only be realized in the mSUGRA scenario in a small parameter
region: $50 \alt \tb \alt 55$, 
$103 \gev \alt \mh, \MA \alt 113 \gev$ and 
$\mH = 115 \pm 2 \gev$, where a significant suppression of
$\sin^2(\beta - \aeff)$ (i.e.\ the $ZZh$ coupling) occurs. 
%We obtain upper bounds on $\mh$ of 124, 119 and 122 $\gev$ and $\tb$ is 
%excluded up to $\tb\agt{3.3},\ {4.6},\ 3.2$ for the three 
%different scenarios, respectively. 
Table~\ref{P3-05tab2} shows the upper bound on $\mh$ and the exclusion
limit on $\tb$ for the three different scenarios. 
The upper bound on $\mh$ is lower than the one in the
unconstrained MSSM~\cite{mssmhiggs,mhiggslong}, 
and the exclusion limit on $\tb$ is also more 
restrictive. This is caused by the fact that not all parameter
combinations of the unconstrained MSSM can be realized in the three 
SUSY-breaking scenarios that we discussed here.
\begin{table}
\caption{Upper bound on $\mh$ and exclusion limit on $\tb$ 
obtained from case (I) for three different scenarios. Corresponding values
for the unconstrained MSSM are also shown for the purpose of
comparison, where the $\tb$ exclusion limit is obtained in the combination 
of $\mhmax$ and no-mixing scenarios~\cite{mssmhiggs,mhiggslong}.}
\label{P3-05tab2}
\begin{tabular}{rlrlrlrlrl}\hline
\multicolumn{2}{c}{mSUGRA}&\multicolumn{2}{c}{mGMSB}
&\multicolumn{2}{c}{mAMSB}&\multicolumn{4}{c}{unconstrained MSSM}\\ \hline
$\mhmax$&$\alt 124~{\rm GeV}$
&$\ \ \mhmax$&$\alt 119~{\rm GeV}\ \ $
&$\mhmax$&$\alt 122~{\rm GeV}$
&\multicolumn{4}{c}
{$\mhmax\alt 130~{\rm GeV}$}\\
$\tb$&$\agt 3.3$&$\tb$&$\agt 4.6$&$\tb$&$\agt 3.2$&$\tb$&$\agt 2.4$,&
$0.7$&$\agt \tb$ \\ \hline
\end{tabular}
\end{table}

In the mSUGRA scenario, cases~(II) and~(III) 
result in similar allowed regions of parameter space.
While for $M_0$ the whole
range up to $1 \tev$ is allowed, $M_{1/2}$ is restricted to
$M_{1/2} \alt 650 \gev$ and $|A_0|$ is restricted to $|A_0| \alt 2 M_0$. 
A significant enhancement of the $hb \bar b$ coupling due to 
large values of $\sin^2\aeff/\cos^2\beta$ is possible over a
wide range of the mSUGRA parameter space.
Values of $\sin^2(\beta -\aeff) \ll 1$
are always correlated in the mSUGRA scenario with negative values 
of $\Delta m_b$ in the range of $-0.2\alt\Delta m_b\alt{-0.4}$, 
giving rise to an enhancement of the $hb\bar b$ coupling.  
Positive values for $\Delta m_b$ 
are only possible if the Higgs boson couples with
full strength to $W$ and $Z$ (see \cite{asbs1} for details.).

Concerning the underlying mGMSB parameters, 
$M_{\rm mess}, N_{\rm mess}$ and $\Lambda$, no severe restrictions can be
deduced for the cases~(I) and~(II).  
The region of low $M_{\rm mess}$
and low $\Lambda$ are excluded by the experimental and theoretical 
constraints
imposed in our analysis.
Lower values of $N_{\rm mess}$ correspond to higher
values of $\Lambda$, and we only find
allowed parameter regions for $N_{\rm mess} \leq 7$.
In contrast to the mSUGRA case, no values of $\sin^2\aeff/\cos^2\beta < 1$
exist.  In particular, a significant enhancement of the $hb\bar b$ coupling is
possible in the region of the highest values
of $\tb$.
The absolute value of $\Delta m_b$ is smaller in
the mGMSB scenario than in the mSUGRA case and does not exceed 
$|\Delta m_b| = 0.2$. Values of $|\Delta m_b| > 0.1$ are only realized for
$\tb \agt 35$.

In the mAMSB scenario, the experimental and theoretical constraints
imposed in our analysis affect in particular the region of large $m_0$
and $m_{\rm aux}$. We find no allowed models with $m_{\rm aux} \agt 70$~TeV,
mostly due to the imposed naturalness bound.
Concerning the $hb\bar b$ coupling, $\sin^2\aeff/\cos^2\beta$ is always larger
than 0.9, and values of $\sin^2\aeff/\cos^2\beta > 10$ are possible for
large $\tb$.  
Positive values of $\Delta m_b$, are bounded from
above by $\Delta m_b \alt 0.5$. On the other hand, we obtain negative
contributions as large as $\Delta m_b \approx - 0.8$, giving rise to a
strongly enhanced $hb\bar b$ Yukawa coupling.

We also studied the mass spectra in the three soft SUSY-breaking
scenarios assuming that the LEP excess is due to the production of the
$h$ or $H$ boson in the MSSM (cases (II) and (III)), which are shown in 
Table.~\ref{P3-05tab3}.  At Run II of Tevatron, 
neutralino and chargino searches
would be sensitive to part of the parameter space.
For the  LHC, on the other hand, the third generation squarks and gluinos
can always be produced.  
A LC with $\sqrt{s}\alt{1}\tev$ will offer a good opportunity to observe
part of the gaugino and slepton spectra.  
\begin{table}
\caption{Super particle mass 
spectra for Case (II) and 
(III) in the mSUGRA, mGMSB and mAMSB scenarios.}
\label{P3-05tab3}
%\begin{tabular}{crclrclrcrcl}\hline
\begin{tabular}{crclrcrcl}\hline

%mSUGRA&$100\gev\alt$&$M_{1/2}$&$\alt{650}\gev$&
mSUGRA&
$50\gev\alt$&$m_{\tilde\chi_1^0}$&$\alt{300}\gev$&
$\ \ \ 150\gev\alt$&$\mste\ \ \ $&$300\gev\alt$&$\mgl$&\\

%&&$|A_0|$&$\alt 2 M_0$&
&
$100\gev\alt$&$m_{\tilde\chi_2^0,\tilde\chi_1^{\pm} }$&$\alt{550}\gev$&
$\ \ \ 450\gev\alt$&$\mstz\ \ \ $&$100\gev\alt$&$m_{\tilde\tau_1}$&\\

%&&&&
&
$250\gev\alt$&$m_{\tilde\chi_2^{\pm}}$&&
$\ \ \ 300\gev\alt$&$\msbe\ \ \ $&$200\gev\alt$&$m_{\tilde\tau_2}$&\\

%mGMSB&$13\tev\alt$&$\Lambda$&&
mGMSB&
$100\gev\alt$&$m_{\tilde\chi_1^0}$&$\alt{350}\gev$&
$\ \ \ 600\gev\alt$&$\mste\ \ \ $&$600\gev\alt$&$\mgl$&
\\

%&&$N_{\rm mess}$&$\leq 7$&
&
$200\gev\alt$&$m_{\tilde\chi_2^0, \tilde\chi_1^{\pm}}$&$\alt{650}\gev$&
$\ \ \ 700\gev\alt$&$\mstz\ \ \ $
&$100\gev\alt$&$m_{\tilde\tau_1}$&$\alt{400}\gev$
\\

%&&&&
&
$350\gev\alt$&$m_{\tilde\chi_2^{\pm}}$&&
$\ \ \ 650\gev\alt$&$\msbe\ \ \ $
&$200\gev\alt$&$m_{\tilde\tau_2}$&$\alt{600}\gev$
\\

%mAMSB&$20\tev\alt$&$m_{\rm aux}$&$\alt{40}\tev$&
mAMSB&
$50\gev\alt$&$m_{\tilde\chi_1^0, \tilde\chi_1^{\pm}}$&$\alt{200}\gev$&
$\ \ \ 400\gev\alt$&$\mste\ \ \ $&$500\gev\alt$&$\mgl$&
\\

%&$150\gev\alt$&$m_0$&$\alt{1600}\gev$&
&
$200\gev\alt$&$m_{\tilde\chi_2^0}$&$\alt{550}\gev$&
$\ \ \ 600\gev\alt$&$\mstz\ \ \ $&$100\gev\alt$&$m_{\tilde\tau_1}$&
\\ 

%&&&&
&
$250\gev\alt$&$m_{\tilde\chi_2^{\pm}}$&&
$\ \ \ 400\gev\alt$&$\msbe\ \ \ $&$200\gev\alt$&$m_{\tilde\tau_2}$& \\
\hline
\end{tabular}
\end{table}

\section{Conclusion}
We have discussed the implication of Higgs searches at LEP2 
in the scenarios of mSUGRA, mGMSB and mAMSB.
We have found upper bounds on $\mh$, exclusion limits on $\tb$ and 
constraints on the parameter space of the  three different scenarios. 
For case (II) and (III), we furthermore investigated the corresponding
spectra of the SUSY particles in view of future SUSY searches. 

\begin{acknowledgments}
It is pleasure to thank S. Ambrosanio for his collaboration
and the contributions to the works reviewed here.  
S.S.\ has been supported by the DOE grant DE-FG03-92-ER-40701. 
A.D. would like to acknowledge financial support from the
Network RTN European Program HPRN-CT-2000-00148
``Physics Across the Present Energy Frontier: Probing the Origin of
Mass''.
\end{acknowledgments}

\end{document}